\documentclass[12pt,preprint]{emulateapj}
\shorttitle{QSLs and AR outflows}
 
\shortauthors{Baker et al.}

\begin{document}

\title{Magnetic reconnection along quasi-separatrix layers as a driver of ubiquitous active region outflows}

\author{D. Baker\altaffilmark{1}, L. van Driel-Gesztelyi\altaffilmark{1,2,3}, C.~H. Mandrini\altaffilmark{4}, P.~ D\'emoulin\altaffilmark{2}, M.~J. Murray\altaffilmark{1}}
\altaffiltext{1}{University College London, Mullard Space Science Laboratory, Holmbury St Mary, Dorking, Surrey RH5 6NT, UK}
\altaffiltext{2}{Observatoire de Paris, LESIA, UMR 8109 (CNRS), Meudon-Principal Cedex, France}
\altaffiltext{3}{Konkoly Observatory, Budapest, Hungary}
\altaffiltext{4}{Instituto de Astronom\'\i a y F\'\i sica del Espacio, CONICET-UBA, 
CC. 67, Suc. 28, 1428 Buenos Aires, Argentina}

\begin{abstract}
 \emph{Hinode's} EUV Imaging Spectrometer (EIS) has discovered ubiquitous outflows of a few to 50 km s$^{-1}$ from active regions (ARs).  These outflows are most prominent at the AR boundary and appear over monopolar magnetic areas.  They are linked to strong non-thermal line broadening and are stronger in hotter EUV lines.   The outflows persist for at least several days.  Using \emph{Hinode} EIS and X-Ray Telescope observations of AR 10942 coupled with magnetic modeling, we demonstrate that the outflows originate from specific locations of the magnetic topology where field lines display strong gradients of magnetic connectivity, namely quasi-separatrix layers (QSLs), or in the limit of infinitely thin QSLs, separatrices.  
 We found the strongest AR outflows to be in the vicinity of QSL sections located over areas of strong magnetic field.  We argue that magnetic reconnection at QSLs separating closed field lines of the AR and either large-scale externally connected or `open' field lines is a viable mechanism for driving AR outflows which are likely sources of the slow solar wind.
\end{abstract}

\keywords{Sun:reconnection --- Sun:active region --- solar wind }
\section{Introduction}
Since its launch on-board the \emph{Hinode} satellite \citep{kosugi07} on 2006 September 23, the EUV Imaging Spectrometer (EIS) \citep
{culhane07} has produced results that have helped to address many of the open questions remaining in solar physics.  In particular, EIS, with its large field of view (FOV) and excellent spectral resolution, has provided us with the opportunity to investigate plasma flows in all solar environments from coronal holes to active regions (ARs).  One of the most intriguing EIS results is the discovery of ubiquitous hot plasma outflows seen in all ARs.  AR outflows are especially important because they are considered to be a possible source of the slow solar wind \citep{sakao07,harra08}.  

\citet{sakao07} reported \emph{Hinode} X-Ray Telescope (XRT) \citep{golub07} observations of continuous outflows from the edge of AR 10942.  The XRT observations were supported by EIS observations of AR blue-shifted flows reported by \citet{doschek07}, \citet{delzanna08}, \citet{harra08}, \citet{hara08}, \citet{doschek08} and \citet{marsch08}.  All authors describe the physical characteristics of the AR outflows and it is very clear that such outflows are distinct from the impulsive plasma flows that result from fast reconnection events such as X-ray jets.  These persistent AR outflows are located in regions of low electron density and low radiance \citep{delzanna08,harra08,doschek08} at the edges or periphery of ARs \citep{sakao07} and over monopolar areas \citep{doschek08}.  They have been observed to persist at nearly the same location from 1.5 to three days \citep{sakao07,doschek08}.  Blue-shifted line of sight velocities for Fe {\sc xii} $195.12~\mathrm{\AA}$ typically range from a few to 50 km s$^{-1}$ \citep{harra08,delzanna08,doschek08} and are faster in hotter coronal emission lines \citep{delzanna08}.  

Whereas red-shifted (cooling) down flows observed in AR closed loops are well understood, to date there is no general consensus for the mechanism(s) driving blue-shifted AR outflows.  Among the mechanisms proposed are `open' magnetic field funnels playing a role in coronal plasma circulation \citep{marsch08}, impulsive heating at the footpoints of AR loops \citep{hara08}, chromospheric evaporation due to reconnection driven by flux emergence and braiding by photospheric motions \citep{delzanna08},
expansion of large-scale reconnecting loops \citep{harra08}, and continual AR expansion \citep{murray09}.  

When viewing the plethora of EIS velocity maps containing ARs, it is quite striking that the outflows appear to occur at locations where magnetic field lines with drastically different connectivities are rooted or meet.  At such locations outflows are concentrated at boundaries that mark the change in magnetic topology from `open' to closed field or appear over a monopolar area between loops connecting to different regions of opposite polarity (see Figures 2 and 4 of \citet{delzanna08}).  Such locations are called separatrices or, in the general case, quasi-separatrix layers (QSLs) \citep{demoulin96}.  In 3D magnetic field configurations, separatrix surfaces separate topological volumes with different magnetic connectivities, while QSLs are defined as thin volumes in which field lines display 
strong gradients of magnetic connectivity (for a recent review, 
see \citet{demoulin07}). When these gradients become infinitely large a QSL
becomes a separatrix.  Specifically, separatrices are present in `open'-closed magnetic topology.  QSLs are preferential locations for current layer development and magnetic reconnection in the absence of magnetic nulls and `bald patch' separatrices 
\citep{demoulin97,milano99,aulanier05,titov08}.  At QSLs, field lines continuously slip across each other during the reconnection process, leading to successive rearrangements of the connections between neighboring field lines along the QSLs, as was shown by MHD simulations \citep{aulanier06} and inferred from \emph{Hinode} XRT observations \citep{aulanier07}.

The relationship between separatrices and QSLs in 3D and observations has been explored in many different solar magnetic configurations in recent years, from small-scale X-ray bright points (XBP) to large-scale X-class flares.  \citet{mandrini96} concluded that the brightness evolution of an XBP was linked to magnetic reconnection at QSLs.  \citet{fletcher01} associated transition region brightenings with both QSLs and `bald patch' separatrices.   

When flares have been studied and related to magnetic reconnection at QSLs, H$\alpha$ and UV flare brightenings were found along or next to QSLs \citep{demoulin97}.  In an X1 flare, \citet{gaizauskas98} reported that plage brightenings and flare kernels were located at the intersection of QSLs with the photosphere.  Flare kernels have been successfully compared to the photospheric and chromospheric traces of QSLs.  In addition, \citet{demoulin97} and \citet{mandrini97} found concentrated electric currents along the boundaries of QSLs where magnetic energy is believed to be stored  in the magnetic field associated with these currents.  The release of free magnetic energy may occur when the thicknesses of QSLs, and their associated current layers, are small enough for reconnection to take place.  \citet{mandrini97} calculated the thickness of a QSL located over a single polarity, where an XBP was observed, to be less than 100 m during the lifetime of the XBP.  Recently, a combination of slip-running reconnection along QSLs before and
after reconnection at a null-point embedded within the QSLs was shown to
explain the observed dynamics of flare ribbons \citep{masson09}.
Theoretical magnetic field topology studies based on QSLs have withstood the test of solar flare observations in the recent past and have confirmed that reconnection is the main physical process in solar flares.  With new instruments such as  \emph{Hinode's} EIS, we look to further test the role of QSLs in association with AR outflows.

In this paper, we propose that the answer to what drives the persistent and ubiquitous AR outflows lies in the AR magnetic field topology.  The AR outflows are observed near or along QSLs.  In the following sections we apply this idea to AR 10942 by computing its magnetic topology and comparing the location of QSLs with outflow regions identified in the EIS data.
 
\section{Data Reduction}

AR 10942 appeared at the Sun's eastern limb on 2007 February 16.  As it crossed the solar central meridian on the 22nd, the AR was measured to have a magnetic flux of approximately 4$\times$10$^{21}$ Mx.  The AR, oriented east-west with a leading negative polarity, was observed by the \emph{Hinode} satellite at various times between February 19 and 26.  Here we concentrate on EIS observations on February 20 and 21. 

No single EIS observation covered the full extent of AR 10942 so multiple data sets and, hence, EIS studies were used to analyze both the eastern and western sections of the AR.  Approximately 24 hours separated the observations.  A raster scan using the $2\arcsec$ slit and consisting of 120 pointing positions with exposure time of five seconds per position was performed with EIS from 23:45 to 23:55 UT on 2007 February 20 (Study ID 37).  The EIS FOV  was $240\arcsec$$\times$$240\arcsec$ and covered most of the AR.  A different raster scan using the $1\arcsec$ slit with exposure time of 30 seconds per position was performed from 11:16 to 11:37 UT on February 20 (Study ID 57).  The FOV was narrower ($41\arcsec$$\times$$400\arcsec$), however, it covered the core eastern outflow region of the AR.  On the western side of the AR, a raster scan with a FOV of $128\arcsec$$\times$$512\arcsec$ included the outflow region that was not fully covered in the raster timed at 23:45 UT on the 20th.  This raster scan using the $1\arcsec$ slit with 60 seconds exposure time ran from 11:40 to 13:48 UT on February 21 (Study ID 45).  EIS Fe {\sc xii} intensity maps of each scan are shown in the left panels of Figures~\ref{fig1}, \ref{fig2}, and \ref{fig3} (23:45 UT and 11:16 UT on the 20th and 11:40 UT on the 21st, respectively).

EIS data reduction was carried out using standard SolarSoft EIS procedures.  Raw data were corrected for dark current, hot, warm and dusty pixels, and cosmic rays.  Relative Doppler velocities were determined by fitting a single Gaussian function to the calibrated spectra in order to obtain the line center for each spectral profile.   A fitted line center was further corrected by removing instrumental effects including slit tilt and orbital variation.  Blue shifts (red shifts) seen in the final velocity maps in the middle panels of Figures~\ref{fig1} to \ref{fig3} correspond to negative (positive) Doppler velocity shifts along the line of sight.  Standard SolarSoft procedures were applied to data from \emph{Hinode} XRT and Michelson Doppler Imager (MDI) \citep
{scherrer95} on-board the \emph{Solar and Heliospheric Observatory} (SOHO).  Additional descriptions of the AR and its surrounding coronal field can be found in \citet{sakao07} and \citet{harra08}.

Co-alignment of EIS data with the underlying photospheric magnetograms, from which magnetic field extrapolations are made and QSLs are calculated, is a crucial step in determining whether the blue-shifted outflows observed at the periphery of the AR are located at or near QSLs.  First, full-disk XRT images were co-aligned with full-disk MDI magnetograms using the solar limb location, after which on-disk features were matched.  Then, EIS Fe {\sc xii}~and Fe {\sc xv}~$284.16~\mathrm{\AA}$ images were co-aligned with XRT images.  MDI magnetic field data were overlaid on all images for final confirmation of alignment.

In the our analysis of AR 10942, we have selected only those data sets which have sufficient photon counts for fitting line profiles.  EIS exposure times ranged from 5 to 60 seconds, thus data quality was affected for some weaker lines.  In addition, MDI data cadence was not ideal with only four magnetograms spanning the EIS data period.  Data quality can affect how well outflows observed in EIS velocity maps match QSL locations as we will discuss in \S4.6. 

\section {Magnetic Field Modeling and Topology}
The Quasi-Separatrix Layers Method (QSLM) and the properties of QSLs have been discussed in detail by \citet{demoulin96} and reviewed by \citet{demoulin06,demoulin07}.  Here we provide only a brief description of the magnetic field modeling technique and the application of the QSLM including recent improvements and some modeling limitations.  We then focus on the specific results obtained for AR 10942.

\subsection{The Magnetic Field Model}
To compute the magnetic field topology of AR 10942, we first need to
model the coronal field. The line of sight magnetic field of AR 10942 is
extrapolated to the corona using the discrete fast Fourier transform
method under the linear force-free field (LFFF) hypothesis ($\vec{\nabla} \times
\vec{B}= \alpha \vec{B}$, where $\vec{B}$ is the magnetic field and
$\alpha$ is a constant). As AR 10942 is not at disk center on February 20 and 21, we do a transformation of coordinates from the observed
to the local frame, as discussed in \citet{demoulin97}.

  We use as the boundary condition for the coronal magnetic model, the MDI magnetogram closest in time to each EIS map. Therefore, since we
have three different EIS scans, we compute three different models.
The value of the free parameter of each model, $\alpha$, is set to
best match the loops observed either by EIS in Fe {\sc xii} or by XRT
depending on whether the EIS FOV is large enough to identify the
global shape of loops.  The procedure we follow is discussed in \citet{green02}.  The best matching values of $\alpha$ are 9.4$\times$10$^{-3}$ Mm$^{-1}$ for the EIS maps starting at 11:16 UT and 23:45 UT on February 20 and 6.3$\times$10$^{-3}$ Mm$^{-1}$
for the EIS map at 11:40 UT on February 21.

In all of our modeling figures, there are a number of field lines which leave the computational box, particularly, those field lines rooted in the vicinity of a QSL.  The LFFF hypothesis is not well suited for modeling `open' field lines because the photospheric magnetic flux is forced to be balanced within the box.  We note the original imbalance in our magnetic field data was
approximately 2.2 G uniformly distributed in a FOV centered at the location of the inner small bipole and extending in 200 Mm in both east-west and north-south directions. 

The fast Fourier-transform method used in our LFFF extrapolations may lead to artifacts due to the periodic nature of the solution.  By enlarging the computational box, we decreased the effect of the periodicity so the weak influence of the box size on the stability of the QSLs locations indicates that the periodicity is not a major issue.  We further tested this using a potential field where there is no intrinsic limitation of the box size ($\propto 1/\alpha$).  The QSLs remained at the same locations as the box was increased in size.  These potential extrapolations were qualitatively compared to the spherical source-surface extrapolations of the same AR in \citet{sakao07}.  
They show field lines originating in the vicinity of the outflows on the eastern side of the AR as we do in Figure~\ref{fig1}, supporting the veracity of the existence of `open' or large-scale field lines in our models.

\subsection{Brief Description of the Quasi-Separatrix Layers Method}

QSLs are defined as regions where there is a drastic change in
field line connectivity  \citep[see e.g.][]{demoulin96}, as opposed
to the extreme case of separatrices where the connectivity is discontinuous. Consider the 
mapping from one photospheric polarity to the opposite
one, denoted by $\vec{r}_{+}(x_{+},y_{+}) \mapsto \vec{r}_{-}(x_{-},y_{-})$, and
the reversed mapping $\vec{r}_{-}(x_{-},y_{-}) \mapsto
\vec{r}_{+}(x_{+},y_{+})$. These mappings can be represented by the vector 
functions $[X_{-}(x_{+},y_{+}), Y_{-}(x_{+},y_{+})]$ and
$[X_{+}(x_{-},y_{-}), Y_{+}(x_{-},y_{-})]$, respectively. 
For example, a QSL is present at $(x_{+},y_{+})$ 
when $X_{-}(x_{+},y_{+})$ and/or $Y_{-}(x_{+},y_{+})$ depend strongly on 
$x_{+}$ and/or $y_{+}$. The strong variation of these functions, $X_{-}$
and/or $Y_{-}$, is found when computing the norm of the connectivity
gradient as described below. The norm
$N(\vec{r}_{+})$ of the Jacobian
matrix in Cartesian coordinates is

\begin{eqnarray} N_{+} &  \equiv &    N(x_{+},y_{+}) 
  %\cr  &   & 
   =  \bigg[       \left( \partial X_{-} \over \partial x_{+} \right)^2
      + \left( \partial X_{-} \over \partial y_{+} \right)^2  \cr 
 &  &      \hspace*{2.2cm}     + \left( \partial Y_{-} \over \partial x_{+} \right)^2
          + \left( \partial Y_{-} \over \partial y_{+} \right)^2   \bigg]^{1/2}  \,.  
  \label{N+}
\end{eqnarray} 

In a similar way, in the negative polarity we have

\begin{eqnarray} N_{-} &  \equiv &    N(x_{-},y_{-}) 
  %\cr  &   & 
   =  \bigg[       \left( \partial X_{+} \over \partial x_{-} \right)^2
      + \left( \partial X_{+} \over \partial y_{-} \right)^2  \cr 
 &  &      \hspace*{2.2cm}     + \left( \partial Y_{+} \over \partial x_{-} \right)^2
          + \left( \partial Y_{+} \over \partial y_{-} \right)^2   \bigg]^{1/2}  \,.  
  \label{N+}
\end{eqnarray} 

%\begin{eqnarray}
 %N_{-} &\equiv &  N(x_{-},y_{-}) 
    %     =   \sqrt{ %\left[
      %  \left( \partial X_{+} \over \partial x_{-} \right)^2
      %+ \left( \partial X_{+} \over \partial y_{-} \right)^2
      %+ \left( \partial Y_{+} \over \partial x_{-} \right)^2
      %+ \left( \partial Y_{+} \over \partial y_{-} \right)^2 } \,.
                 % \right]^{1/2}
  %\label{N-}
%\end{eqnarray}
 
  A QSL was first defined by the condition $N_{+} >> 1$ and $N_{-} >> 1$
in both photospheric polarities \citep{demoulin96}.
However, for a field line linking photospheric locations
$(x_{+},y_{+})$ and $(x_{-},y_{-})$, both of which have different
normal field components $B_{z+}$ and $B_{z-}$, the definition of a
QSL given by Equations~(\ref{N+}) and~(\ref{N-}) implies that $N(x_{+},y_{+}) \neq
N(x_{-},y_{-})$ if $B_{z+} \neq B_{z-}$. \citet{titov02} defined
another function to characterize QSLs which is independent of the
mapping direction, the squashing degree $Q$. It was shown that 
$Q$ can be simply defined by the product of the values of $N$ determined
when starting the mapping of field lines
from both of their photospheric footpoints, therefore
\begin{equation}
 Q \equiv N_{+}  N_{-} \,.
\end{equation}
Then, a QSL is defined when
$Q>>2$; the value $Q=2$ is the lowest possible value. This value is 
found when, for example, $x_{+} = -x_{-}$ and $y_{+} = y_{-}$ (as present in a simple
potential arcade oriented along the $y$ direction).
On the other hand, $Q$ becomes infinitely large when the field line mapping is
discontinuous, {\it i.e.} when we have separatrices. By definition,
$Q$ is uniquely defined along a field line by $(\vec{B} \cdot
\vec{\nabla}) Q = 0$.  

The physical meaning of this new definition
can be explained as follows. If we consider an elementary flux tube
rooted in an infinitesimal circular region with a given polarity
sign, $Q$ measures the aspect ratio of the distorted ellipse defined
by the mapping of this flux tube footpoint in the other polarity
sign. That is, $Q$ measures how much the initial elementary region
is squashed by the mapping.  MHD simulations have shown that the thickness 
of a QSL is related to the current density that develops in it, such that 
the thinner the QSL, the higher the current density \citep{aulanier07}.  More 
specifically, the thickness is defined as the full width at half maximum of 
the $Q$ profile that is computed along a 1D segment that crosses the 
photospheric QSL trace.

$Q$ can be computed only for field lines reaching the lower boundary at both ends (\emph{i.e.} the field lines are closed). A fraction of the lower boundary is magnetically connected to one side or to the top of the computation box.
Such field lines either extend into interplanetary space or they have long connections outside the AR.  We refer to them as `open' field or large-scale loops. In-between truly `open' and closed field lines a separatrix is present.  Moreover, in-between the closed field lines of the AR and large-scale externally connected field lines, a QSL is generally expected since one footpoint of the connection depends drastically on the position of the other footpoint which stays inside the AR.  With our extrapolation procedure, we cannot distinguish between these two
cases and we can only compute the transition between closed and `open' (box-reaching) field lines.  This transition is kept by imposing an arbitrarily high value of $Q$ (larger than the minimum value of $Q$ used in the figures shown).  We discuss how the transition is affected by the size of the computational box in \S{4.2}.

The finite size of our computational box does not allow for us to distinguish between truly `open' and large-scale field lines.  Indeed, some of the field lines leaving our computational box remain `open' in spherical source-surface computations carried by \citet{sakao07} (see Figure 4B).  Other field lines are truly large-scale connecting to a neighboring active region as shown by \citet{harra08} or far quiet Sun regions, therefore, we are confident that the high-$Q$ dominant QSLs are not artifacts resulting from the methodology.

The numerical procedure used to determine the values of $Q$ in
this work has been thoroughly discussed by \citet{aulanier05}. Our
magnetic field model takes observed magnetograms as the boundary condition,
therefore, the presence of parasitic polarities in the
configuration (e.g. see MDI magnetogram at 11:16 UT on February 20 in Figure~\ref{fig2}) results in 
multiple QSLs. However, only those corresponding to the highest values of 
$Q$, in other words, the thinnest QSLs lying on both
main positive and negative AR polarities are considered. 
%The photospheric traces of the QSLs shown in all figures correspond to values of $\log_{10}Q$ above $\approx 10$.  QSLs with lower values of $\log_{10}Q$ are not shown.  

The magnetic models together with the QSL locations are shown in the right panels of Figures 1 to 3.  %This is \textbf{the first time} that the QSLM, using the definition given by Equation (2), is applied to observed magnetic field data.
This is only the second time (see \citet{masson09}),
that the QSLM using the definition given by Equation (2) has been applied to observed magnetic
field data, though there are some differences in the methodology between
our approach and that of \citet{masson09}.

\section{Results}
\subsection{Locations of Dominant QSLs}
Before comparing the locations of the QSLs calculated by the QSLM with the locations of the observed AR outflows, we first consider the correctness of the extrapolations by comparison with coronal observations of AR 10942.  Figure~\ref{fig4}, left panel, shows the large-scale topological structure of the AR and its surroundings based on the MDI magnetogram closest to the EIS observation time of 23:45 UT on February 20.  There is a global agreement between the coronal magnetic field model with the XRT observations in Figure 4 and with the EIS Fe {\sc xii} intensity map in Figure~\ref{fig1}, left panel.  Though not shown, the global structure is similar to the large-scale coronal magnetic field model calculated at the time of each EIS observation, 12 hours earlier on the east side (Figure 2) and 12 hours later on the west side (Figure 3).

Figure~\ref{fig4}, right panel, shows the photospheric trace of the dominant QSLs (indicated by thick red lines) overlaying \emph{SOHO} MDI magnetic field isocontours.  The most extended QSL (labeled as \emph{a}) is located over the following positive polarity of the AR (eastern side) where we see the strongest outflows in the EIS observation on February 20 at 23:45 UT (Figure 1).  A major QSL (labeled as \emph{d}) is found over the leading negative polarity and is associated with outflows visible in the EIS velocity map in Figure~\ref{fig3}.  In the following sections, QSLs with values of $\log_{10}Q$ above
$\approx 10$, will be referred to as dominant QSLs.

\subsection{Stability of QSL Locations}
The limited size of the computation box can influence the extrapolated field, and in particular the limit between closed and `open' or `box-reaching' field lines.
We therefore tested how the locations of our dominant QSLs depend on the computational box size and found the QSL locations on the positive polarity were stable.  The limit between closed and `open' or `box-reaching' field lines was also stable.  The QSL on the negative preceding polarity somewhat decreased in extension, however, the QSL section associated with the core outflows did not change.  Figure~\ref{fig5} shows a side-on view of the enlarged computational box and the resulting photospheric traces of the dominant QSLs overlaid on the same large FOV EIS velocity map from Figure~\ref{fig1}.  Dominant QSLs (Figure~\ref{fig5}, right panel) are similar to those shown in Figure~\ref{fig4} before the computational box was enlarged.

\subsection{Fe {\sc xii} Flows}
In the velocity map of the large FOV (Figure~\ref{fig1}, middle panel), there is a series of loop structures connecting the positive and negative magnetic field concentrations of the AR.  These loops are red-shifted, indicating downflows within the loop structures.  Line of sight downflow velocities range from a few km s$^{-1}$ up to a maximum of 32 km s$^{-1}$.  The eastern and western region velocity maps shown in Figures~\ref{fig2} and \ref{fig3}, respectively, are dominated by outflows with small patches of downflows up to 14 km s$^{-1}$.
 
Blue-shifted outflow regions are observed at the periphery of the AR in all velocity maps.  In addition, outflows are located over the monopolar magnetic field concentrations of the AR.  Outflows are strongest over each of the strongest magnetic field concentrations, especially to the east.  This is shown in the middle panels of Figures~\ref{fig1} to \ref{fig3} where MDI magnetic field isocontours of $\pm$ 50 G are overlaid on each velocity map. Maximum line of sight velocity for the eastern side of the AR is -49 km s$^{-1}$ in the raster scan at 23:45 UT on February 20.     Outflows observed in the maps starting at 11:16 UT on the 20th and at 11:40 UT on the 21st have velocities of -14 and -13 km s$^{-1}$, respectively.  These values and properties of the AR outflows are consistent with analysis carried out by previous authors studying this particular AR.

\subsection{Relationship between Fe {\sc xii} Outflows and QSLs}
\begin{figure}[h]
\epsscale{1.0}
\plotone{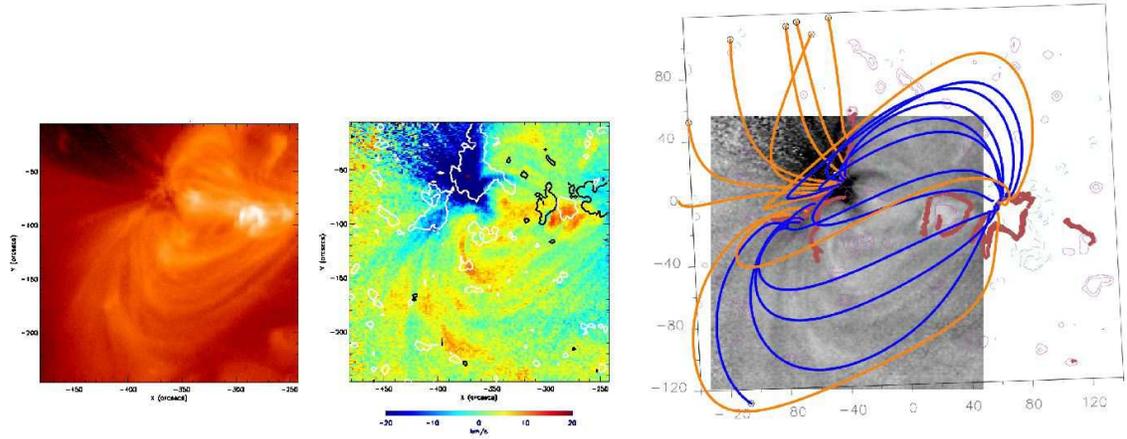}
\caption{Left panel - EIS Fe {\sc xii} emission line intensity map of AR 10942 at 23:45-23:55 UT on 2007 February 20.  Middle panel - EIS Fe {\sc xii} emission line velocity map overlaid with $\pm$ 50 G MDI magnetic contours.  White/black is positive/negative polarity.  Right panel - photospheric trace of QSLs (thick red lines) and field lines originating in the QSLs are overlaid on a grayscale EIS Fe {\sc xii} emission line velocity map.  Orange/blue field lines are drawn from the western/eastern side of the eastern QSL over the positive polarity and lines with circles leave the computational box and are considered to be `open' or large extended loops. The coordinate system is centered on the AR instead of the Sun and both axes have units of Mm.  Magnetic field isocontours are shown in continuous pink/dashed blue lines for positive/negative values of the field ($\pm$ 20, $\pm$ 50, and $\pm$ 500 G).    The overlay image clearly shows strong AR outflows along `open' field lines computed from the eastern side of the QSL located over the positive polarity.  Note, the size of the computational box for this figure and Figures 2, 3, 4, and 6 is
400 Mm in all directions.  \label{fig1}}
\end{figure}

\begin{figure}
\epsscale{1.0}
\plotone{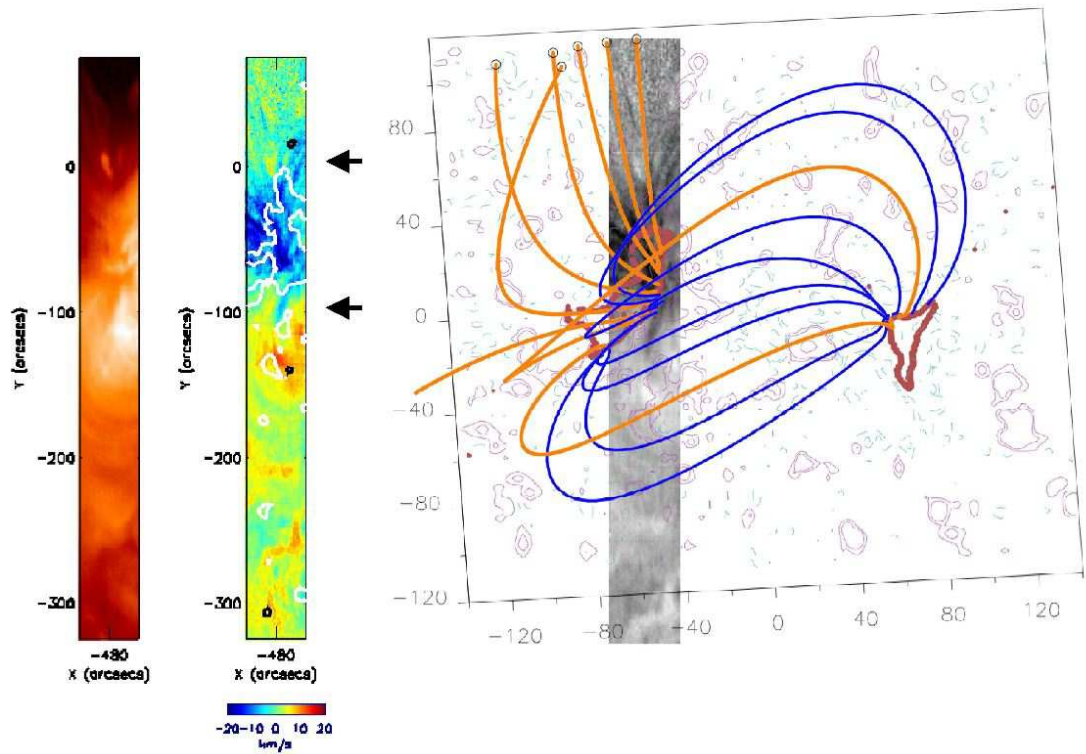}
\caption{EIS Fe {\sc xii} emission line intensity (left panel) and velocity (middle panel) maps and photospheric trace of QSLs and field lines originating in the QSLs (right panel) at 11:16-11:37 UT on 2007 February 20.  The drawing convention is the same as that used in Figure~\ref{fig1}.  The image in the right panel shows AR outflows along `open' field lines computed from the eastern side of the QSL.  (Black arrows indicate the zoomed FOV shown in Figure~\ref{fig6}).  \label{fig2}}

\end{figure}

\begin{figure}
\epsscale{1.0}
\plotone{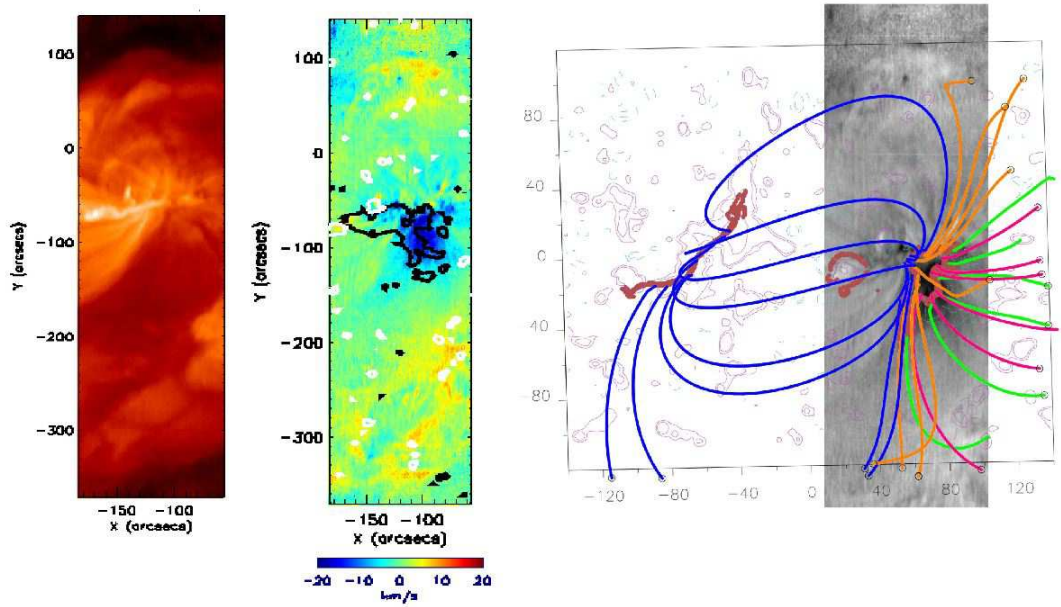}
\caption{EIS Fe {\sc xii} emission line intensity (left panel) and velocity (middle panel) maps and photospheric trace of QSLs and field lines originating in the QSLs (right panel) at 11:40-13:48 UT on 2007 February 21.  The drawing convention is the same as that used in Figure~\ref{fig1}, with the addition of green/pink field lines computed from the east/west side of the western part of the QSL trace.  The AR negative polarity is connected to the AR positive polarity and to the positive polarity of a neighboring bipole to the west.  The overlay image shows AR outflows along `open' field lines computed from the inner side of the closed QSL trace. { Note that the MDI magnetic map used as boundary condition for the modeling was taken at 08:03 UT, 3.5 hours prior to the start of the EIS scan, due to patchy magnetic data coverage. } \label{fig3}}
\end{figure}

The right panels in Figures~\ref{fig1} to \ref{fig3} illustrate the relationship between QSLs and the blue-shifted outflows observed in EIS Fe {\sc xii} velocity maps of both the eastern and western main AR polarities.  Across all of these panels, the outflows consistently occur in the vicinity of QSLs.  Unlike QSLs and flare kernels, the relationship is more subtle since kernels are formed in a relatively thin layer of the atmosphere whereas outflows are observed over a broad range of coronal heights.  A direct comparison for coronal flows is further complicated by the fact that our 2D velocity maps result from the integrated, optically thin emission along the line of sight over a large depth.  Presently, it is not possible to deconvolve these 2D maps in order to obtain the 3D locations of the observed flows.  Moreover, for an AR observed away from the solar central meridian, the line of sight integration is significantly different from integration along the local vertical, thus creating projection effects.  

Though we cannot determine the precise 3D structure of the velocities, 
we can compare the projection of the expected locations of the outflows 
with the observed velocity maps.  Outflows are expected in the vicinity of QSLs where reconnection can transform closed loops into `open' field or large-scale loops.  The dense plasma of the initial closed loop is no longer confined along the reconnected magnetic field and is accelerated by a plasma pressure gradient and a magnetic tension force.  Then, as in previous flare studies, we show both the photospheric trace of QSLs and the field lines rooted on both sides of QSLs.  For velocities observed in hot lines (e.g. Fe {\sc x} $184.54~\mathrm{\AA}$ and Fe {\sc xii}), it is the spatial extension of the `open'/large-scale field lines which is the most relevant to compare with the spatial distribution of the observed outflows. Indeed, we found that such a set of field lines greatly spreads out from the QSL photospheric trace and they fill a spatial region which is comparable to the observed outflows, as explained below.

Figure~\ref{fig1} shows the largest map obtained of this AR. Two QSLs are present on the main positive (following) polarity at this time (see Figure~\ref{fig4} for a better view of the photospheric trace of the QSLs). In the right panel of Figure~\ref{fig1}, orange (blue) colored field lines have been computed with integration starting on the east (west) side of the QSL labeled as \emph{a} in Figure~\ref{fig4}.  Field lines ending in a black circle have reached the computational box and are considered to be `open' or large-scale loops. These `open' field lines are found to overlay relatively well the observed strongly blue-shifted outflows.  Since projection effects are taken into account in the magnetic model, this indicates that the strong outflows are plausibly coming from the vicinity of this QSL.

Figure~\ref{fig2} shows the magnetic field structure and outflows associated with the eastern (following) polarity of the AR, $\approx{12}$ hours before that shown in Figure~\ref{fig1}. 
At that time, we find that the shape of the photospheric trace of the QSL over the positive polarity is closed rather than the two open QSL traces found in Figure~\ref{fig1} and labeled as \emph{a} and \emph{b} in Figure~\ref{fig4}. However, this difference is not important for the present study since the extension of the QSL trace shown depends on the choice of the minimum value of $\log_{10}Q$, {\it i.e.} a QSL trace can appear to be closed or not depending on this minimum value.  EIS outflows are associated with the westernmost section of the closed QSL trace from where we have computed field lines starting from both sides (Figure~\ref{fig2}). This is basically the same configuration shown in Figure~\ref{fig1}.   On the easternmost section of the closed QSL trace, a drastic change of connectivity is also present between `open'/large-scale field lines and small-scale ones connecting small negative polarities on the east side of the AR.  We do not show these field lines so that Figure~\ref{fig2} is not overcrowded, however, similar connectivities with a mirrored symmetry are shown in Figure~\ref{fig3} at the westernmost side of the AR.  
 
 On the western side of the AR shown in Figure~\ref{fig3} the photospheric trace of the QSL is similar to that labeled as \emph{d} in Figure~\ref{fig4}.  For this observation, we show field lines all around the QSL using a different color for field lines computed from the 
eastern and western QSL sections (Figure~\ref{fig3}). The western side has 
similar connectivities to those of the eastern side of the following 
polarity, {\it i.e.} the QSL separates short field lines 
(green, connecting the 
main polarity to network-like polarities) from `open' ones (pink). 
Similarly, the blue and orange field lines are analogous to those found in 
Figures~\ref{fig1} and~\ref{fig2} for the following polarity.  Again, the 
blue shifts are mainly found along `open' field lines located in the 
vicinity of the QSL.  

\subsection{Relationship between Si {\sc vii} Outflows and QSLs}
To complement the EIS Fe {\sc xii} observation in Figure 2, data in the cooler Si {\sc vii} $ 275.35~\mathrm{\AA}$ spectral line with a signal to noise ratio large enough to 
detect well the velocities over the AR are analyzed. This allows a closer comparison of 
outflows with the calculated QSLs as the emission comes from a less spatially 
extended region.
We are able to determine which section of the QSL is related to the strong coronal outflows.  In Figure~\ref{fig6}, zoomed velocity maps from two cooler emission lines, Si {\sc vii} and Fe {\sc x} ($\log_{10}T_{max}$ = 
5.8 and 6.0, respectively), are shown with the corresponding zoomed EIS Fe {\sc xii} map ($\log_{10}T_{max}$  = 6.1).  All maps are overlaid with the positive polarity magnetic contours (white = 100 G and red/blue = 500 G) and the (black) photospheric trace of QSL (labeled `a' and `b' in Figure~\ref{fig4}) at 11:16 - 11:37 UT on February 20 (c.f. Figure~\ref{fig2}, right 
panel).  Figure~\ref{fig6} provides evidence that the strongest
outflows occur in the vicinity of the strongest magnetic field
concentrations along the dominant QSLs with values of $\log_{10}Q$ above
$\approx 10$.

The EIS Si {\sc vii} velocity map is also shown alone (Figure~\ref{fig6}, panel A) so that the 
blue outflow lanes (marked by arrows) can be better seen.  These weak blue outflow 
regions are narrow and elongated, as would be expected if the outflows are a 
result of reconnection along a QSL or separatrix.  Indeed, these narrow blue regions lie 
close to the western part of the closed QSL and appear to be the base of the 
extended blue shifts seen fanning out in the velocity maps of the hotter 
Fe {\sc x} and Fe {\sc xii} emission lines.  We make this observation with the caveat that it is often too tempting to consider outflows at different temperatures to be stratified by height in the solar atmosphere.   There is a slight difference in position 
between the QSL and the northern narrow blue-shifed region observed in 
Si {\sc vii}  (Figure~\ref{fig6}, panel A), though the core of the strongest outflows does lie close to the strongest magnetic field and a small section of the western QSL.   Slight differences in position of the QSL are most likely due to the fact that our LFFF model includes only a global magnetic shear through a unique value of 
$\alpha$.  Typically, $\alpha$ is non-uniform in ARs and a significant 
departure from the mean value is found in vector magnetograms.  Furthermore, due to unusually scarce MDI magnetic field maps coverage, the magnetic map used as a boundary condition for the modeling was taken at 08:03 UT, 3.5 hours prior to the start of the EIS scan, and magnetic evolution during this period may result in some differences between the computed QSL and observed flow locations. 
 
The Si {\sc vii} velocity map  (Figure~\ref{fig6}, panels A and B) is dominated by red-shifted downflows, showing strong resemblance to the pattern of upflows observed in the higher-temperature Fe lines (Figure~\ref{fig6}, panels C, D), while being not exactly co-spatial with them.  Stronger downflow lanes (deep red in Figure~\ref{fig6}, panels A and B) are separated by weak downflow lanes (green) and by two upflows lanes (indicated by the two arrows).  In Figure~\ref{fig6}, (panel E), a contour (white) of Si {\sc vii} downflows (= 5 km s$^{-1}$) is overlaid on the Fe {\sc xii} velocity map (the offset between the two EIS CCDs has been taken into account).  The contour is consistent with the outline pattern of the hotter upflows. The red-shifted loop-like features (Figure~\ref{fig6}, panels A and B), which are visually better defined than the blueshifted structures in the hot Fe {\sc x} and {\sc xii} lines, appear to converge towards the QSL, most of them ending on its western side.  The pattern of the downflow structures in Si {\sc vii} `trace' the slightly displaced outflows in the hotter Fe lines.  Since QSLs indicate regions where reconnection can transform closed loops into
% $`openÕ$ Þeld 
`open' field or large-scale loops, we suggest that the red-shifted structures in Si {\sc vii} represent cooling downflows of previous outflows resulting from earlier reconnection events, and thus they provide further evidence that the outflows originate from the vicinity of QSLs and fan out with height. 

\subsection{Do we observe outflows over all QSLs?}
No.  Outflows are {\it not} observed over all QSLs.  We have already seen this on the 
following polarity for the eastern part of the QSL (Figure~\ref{fig6}).  In 
order to drive outflows we need a QSL separating `open' 
or large-scale field lines from closed ones. But this alone is not sufficient.  An evolution of the 
magnetic configuration is needed to first build up significant currents along 
the QSL and the current layer thickness must become small enough 
to induce magnetic reconnection. Conversely, not all QSLs can drive 
outflows, even with the just mentioned field evolution, since the presence of 
large-scale or `open' field lines is required only on  
one side of the QSL. QSLs are typically present inside ARs (see the references on flares in \S1)
and are present in the case of AR 10942 as shown in Figure~\ref{fig7}.  No significant upflows are 
associated with this internal QSL.

Reconnection at QSLs 
in closed, small-scale loops can drive siphon flows by an asymmetric 
deposition of energy in the reconnected loops.  So closed loops can have 
upflows dominant in one leg of the reconnected loops. However, such upflows 
in coronal lines are expected to be mixed up with downflows from other loops,
in particular, those coming after a heating episode when coronal loops are 
cooling down.  Since many heating processes are expected to occur in the 
neighboring closed loops, without any significant phase synchronisation, 
the downflows are likely to be mixed with upflows, so that no clear 
upflow pattern is observed in closed loops outside flaring times (Figures~\ref{fig1}-\ref{fig3}, \ref{fig6},
and references to 
spectroscopic studies, in particular using EIS results, \S 1).

Additionally, it is possible that flows will not be instigated along the whole length of a QSL.  Work done by \citet{demoulin97} on flare ribbon-QSL association showed that the presence of a QSL is not sufficient by itself for flare activity.  As we stated previously in this section, the evolution of the magnetic field must be great enough to build intense current layers (e.g. via twist or shear) which become thin enough for reconnection to take place.  A strong magnetic field is also required to provide enough magnetic energy.  Moreover, not all QSLs are in the appropriate state to become flare-active (e.g. thin enough to reconnect).  We believe this is also a plausible scenario in the context of QSLs and AR outflows.  The energy required would be less but an evolving and strong magnetic field are likely conditions for sections of a QSL with a high $Q$ to become flow-active.  The resulting outflows are a direct consequence of the reconnection.  We can confirm that the strongest flows occur in the vicinity of the QSL sections that overlie the strongest magnetic field (see Figure~\ref{fig6}), however, due to less than ideal data coverage, we were unable to properly study the evolution of the QSLs and, therefore, determine with more certainty how and why certain sections of the QSLs become flow-active.  This is the basis for future work with a wider sample set and better data coverage. 

\section{Discussion and Conclusions}

In this paper, we have used \emph{Hinode} EIS and XRT observations of AR 10942 coupled with magnetic field modeling to analyze and explain AR outflows.  QSL locations were computed from observed magnetic data using the new definition given by \citet{titov02}.  
The strongest outflows are associated with portions of QSLs located over regions of strong magnetic field.  The area and velocity of the outflows increase with temperature. We found that a narrow blue-shifted outflow lane is present along some QSLs in the lowest temperature Si {\sc vii} EIS velocity map where the exposure time is sufficient to have a significant velocity signal to noise level. The outflow area is larger in hotter Fe {\sc x} and even larger in Fe {\sc xii}, indicating that the outflows fan out and accelerate with height (Figure~\ref{fig6}).  Since the hot outflows are not well defined areas, determining their origin is non-trivial. The base of the blueshifted outflows were further constrained by redshifted downflows seen in the Si {\sc vii} velocity map bearing strong resemblance to the pattern of upflows observed in the higher-temperature Fe lines, while being not entirely co-spatial with them. We interpreted these red-shifted structures as cooling plasma flows along loops of previous hot upflows. Since the redshifted loop-like features appeared to converge towards the QSL, most of them ending in its vicinity, they provided further evidence that the outflows do originate from the vicinity QSLs and fan out with height.

The eastern outflow region of AR 10942 was 25-32 degrees in distance from the central meridian leading to projection effects especially in the higher temperature loops along which outflows are 
fanning out from the vicinity of the QSLs. This effect, coupled with an intrinsic optically thin spectral line formation, masks, at least in part, the spatial origin of the outflows. In particular, it is difficult to 
clearly separate two intrinsically different origins such as a shell-like source around QSLs from a volume-like source such as outflows coming from the full `open'/large-scale field region.  The clearest distinction between different spatial origins, so different physical mechanisms, can be best understood from spectral lines formed at the top of the transition region where there are both relatively high velocities and a vertical localization of the line formation.  This requires long exposure times to have a sufficiently high signal to noise ratio. Such data were available only once in the studied AR, however, it was sufficient to confirm the QSL origin of outflows.  

\begin{figure}
\epsscale{1.0}
\plotone{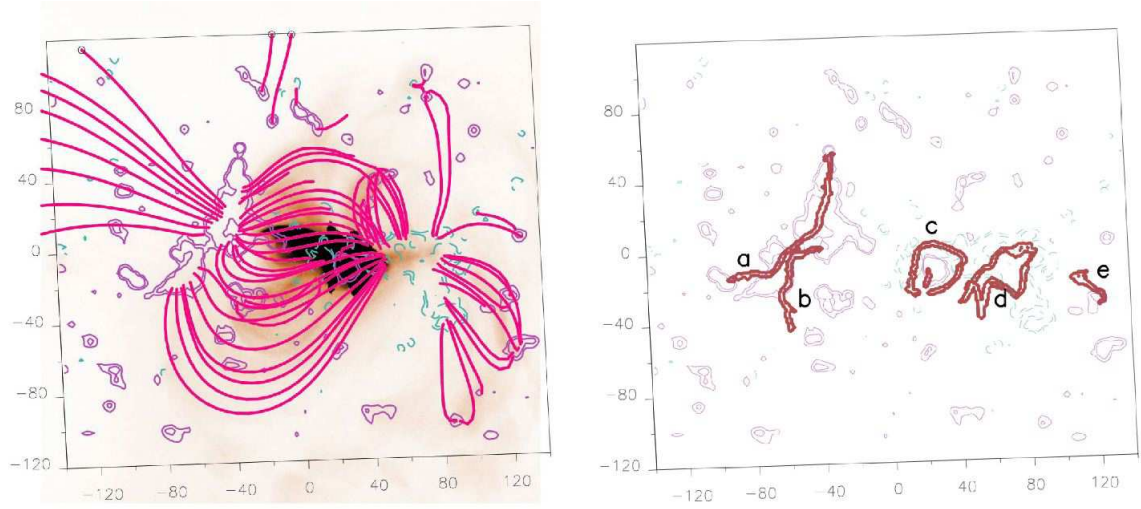}
\caption{Left panel - Global linear force-free magnetic field model of AR 10942 with $\alpha$ = 9.4$\times$10$^{-3}$ Mm$^{-1}$ over \emph{Hinode} XRT image.  There is a global agreement between the coronal magnetic field model with the XRT observations and with the EIS Fe {\sc xii} intensity map in the left panel of Figure~\ref{fig1}.  Right panel - Photospheric trace of dominant QSLs (thick red lines) in AR 10942 with \emph{SOHO} MDI magnetic field contours.   The magnetogram and magnetic model correspond to Figure~1, as does the drawing convention.  The photospheric traces of QSLs have been labeled as \emph{a}, \emph{b}, \emph{c}, \emph{d}, and \emph{e}.  Though the shapes of QSLs and the photospheric field distribution change, we use this labeling to refer to the equivalent QSLs at different times. \label{fig4} }
\end{figure}

\begin{figure}
\epsscale{1.0}
\plotone{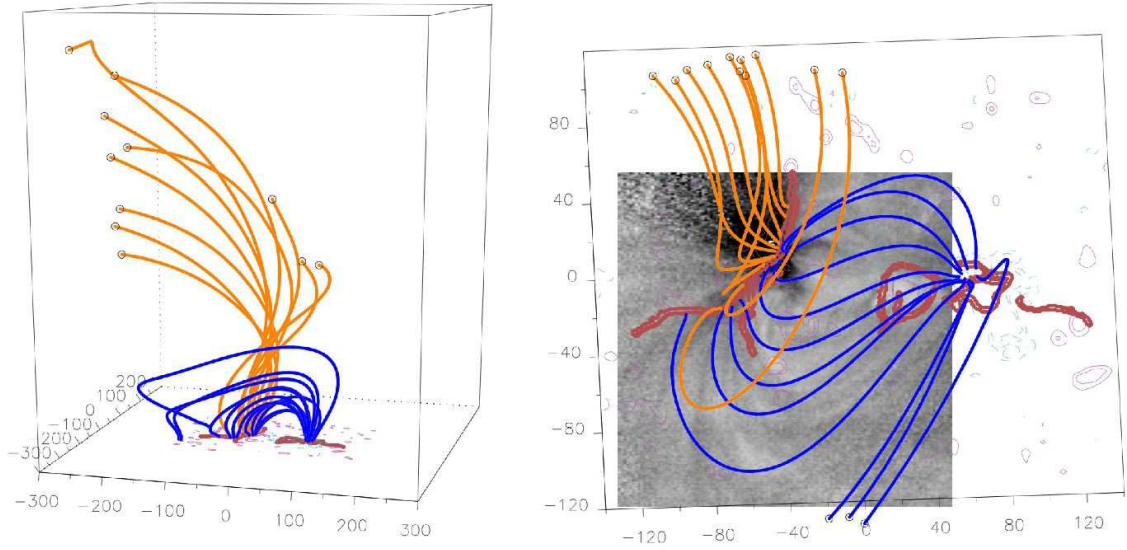}
\caption{Magnetic field model and high-$Q$ QSL locations computed using the same MDI magnetogram as a boundary condition and $\alpha$ as in Figure~\ref{fig1}, but a larger computational box (600 Mm in both east-west and north-south directions and 700 Mm in height).  Drawing conventions are similar to those used in Figures~\ref{fig1} to \ref{fig3}.  Left panel - shows field lines in the full box from an arbitrary point of view.    Orange and blue field lines are drawn from the newly computed QSLs in the same way as in Figure~\ref{fig1}.  Right panel - photospheric trace of QSLs and the same field lines as in the left panel from the observer's point of view overlaid on a grayscale EIS Fe {\sc xii} emission line velocity map (c.f. Figure~\ref{fig1}).  The eastern dominant QSLs (labeled as \emph{a} and \emph{b} in Figure~\ref{fig4}) are wholly stable with the enlargement of the computational box whereas the western QSL (labeled as \emph{d} in Figure~\ref{fig4}) shrinks slightly towards the south.    \label{fig5}}
\end{figure}

\begin{figure}
\epsscale{1.0}
\plotone{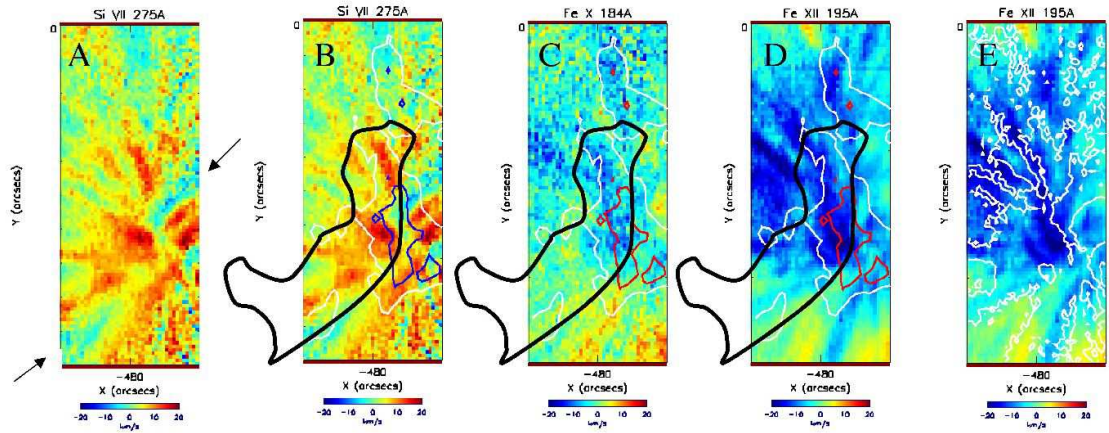}
\caption{Zoomed EIS Si {\sc vii}, Fe {\sc x}, and Fe {\sc xii} emission lines ($\log_{10}T_{max}$  = 5.8, 
6.0, and 6.1, respectively) velocity maps of AR 10942 at 11:16-11:37 UT on 2007 February 20.  Panel A - Si {\sc vii}.  Panel B - Si {\sc vii} overlaid with contours of 100~G (white) and 500~G (blue) magnetic field isocontours.  Panels C and D - Fe {\sc x} and Fe {\sc xii}, respectively, overlaid with contours of 100~G (white) and 500~G (red) magnetic field isocontours.  Thick black contours are photospheric traces of the dominant QSL from Figure~\ref{fig2}, right panel.  Panel E - Fe {\sc xii} overlaid with contours (white) of Si {\sc vii} downflows (5 km s$^{-1}$).  The strongest outflows in the hotter Fe lines occur in the vicinity of the strongest magnetic field concentrations on the western side of the QSL.  Red-shifted downflows evident in Si {\sc vii} appear to `end' on the same side of the QSL.  The pattern of the downflow structures in Si {\sc vii} (panel E) appears to `outline' the slightly displaced outflows in the hotter Fe lines (see panel E).  The narrow outflow lanes in Si {\sc vii} (indicated by black arrows in panel A) appear to be the base of outflow regions fanning out in EIS Fe {\sc x} and 
Fe {\sc xii}  velocity maps (panels C and D).  See \S 4.5 for a detailed discussion of this figure.  \label{fig6}}
\end{figure}

\begin{figure}
\epsscale{1.0}
\plotone{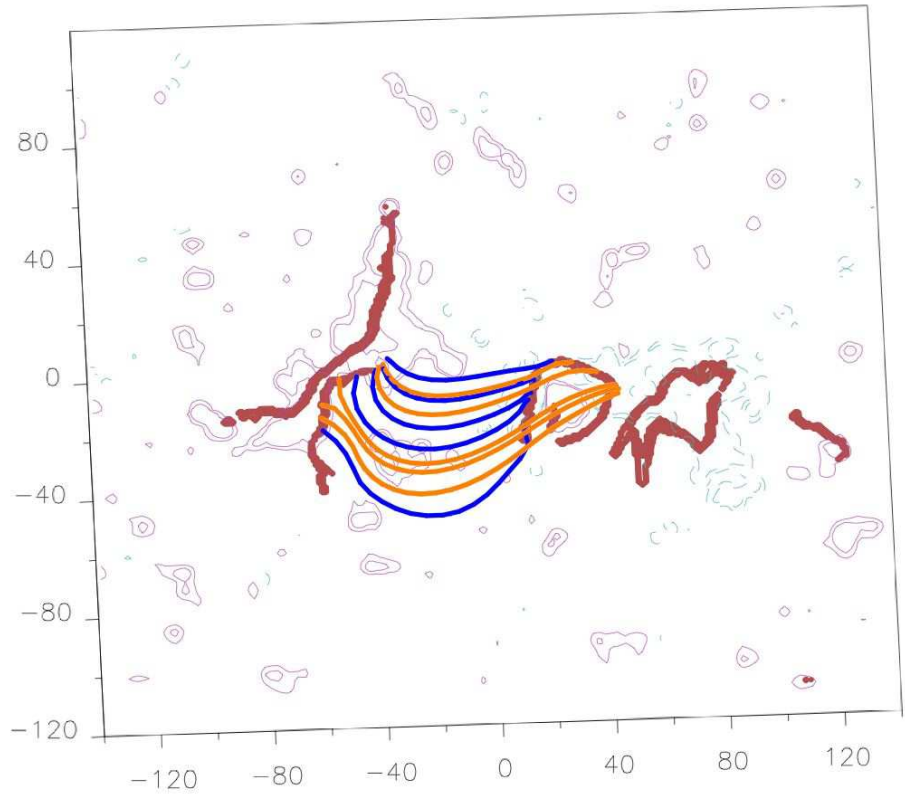}
\caption{Field lines originating from internal QSLs represent low lying loops.  Orange/blue field lines are drawn from the eastern/western side of the QSL trace located on the positive magnetic polarity.  The magnetogram, magnetic model, and drawing convention correspond to Figure 1.\label{fig7}}
\end{figure}

Since QSLs have distinctive characteristics in coronal images, for example clearly separating loops indicating a change in connectivity, we verified visually in EIS maps from other publications (\citet{hara08}, \citet{marsch08}, \citet{doschek08}, \citet{murray09}) the general validity of the relationship between QSLs and AR outflows.  Figure~\ref{fig4} in \citet{delzanna08} shows the clearest example observed so far with multi-line EIS velocity maps of an AR observed close to Sun center so that there are practically no projection effects.  The strongest blue shifts are seen along narrow lanes separating closed AR loops from large-scale loops which appear `open' or connect to distant magnetic polarities.  Outflow regions in this AR demonstrably increase in strength and breadth with temperature and height, similar to what we observe in AR 10942.

There have been different proposals for the mechanisms driving the outflows and nearly all of them published so far fit within the QSL scenario.  The major reasons why AR outflows fit well with the QSL scenario are as follows:
\begin{itemize}
\item QSLs (including separatrices) naturally explain the most puzzling characteristic of outflow regions which is their occurrence over monopolar areas \citep{doschek08}.  Our computations show that QSL locations over the AR polarities are in good agreement with the outflow regions.  By definition, QSLs divide drastically different connectivities over a magnetic polarity (see any of the papers analyzing flare observations cited in \S1).  
\item The strongest outflows are seen at the periphery of ARs.  As suggested by \citet{marsch04, marsch08}, the sharp boundaries found in Doppler velocity maps between blue and red-shifted features mark the change in magnetic topology from `open' or large-scale to closed field.  
\item Another enigmatic characteristic of the AR outflows discovered by EIS is their longevity.  Outflows persist at approximately the same locations for time scales of at least several days \citep{sakao07,doschek08}.  The longevity of the outflows can be explained by the very nature of QSLs.  They are defined by the global properties of the magnetic configuration which evolves slowly.  More precisely,  they are dominantly defined by the photospheric magnetic flux distribution with the exception of highly sheared configurations occurring in the core of ARs and related to large flares (see the reviews by \citet{demoulin06, demoulin07} and references therein).
\end{itemize}

QSLs are locations where ideal MHD breaks down and reconnection takes place.  This reconnection is rarely fast unless the current layer thickness is small enough and/or there is a strong driving force such as an ideal instability of the magnetic field.  
In 3D, reconnection occurs simultaneously at multiple locations along the length of a QSL involving many field lines over an extended area \citep{aulanier06,aulanier07,parnell09}, so outflows do not appear intermittent and patchy but rather smooth and extended.  
Of course, small-scale events with low reconnection rates are also expected to happen along QSLs, especially where they are initially broad.  In these cases the accumulation of magnetic stress is needed to build a thin enough current layer to later start reconnection impulsively with a sufficiently fast rate.  We suggest that the reconnection-driven plasma flows observed on one side of a QSL are the result of the spatial and temporal superposition of nearly-continuous reconnection together with many small-scale events.  These reconnections are driven by the almost permanent shuffling of footpoints. 

Reconnection and consequent energy release lead to particle acceleration followed by heating of plasma.  The heating of plasma occurs through gentle chromospheric evaporation.  Acceleration of particles results in enhanced non-thermal line broadening.  \citet{doschek07}, \citet{delzanna08}, \citet{hara08} and \citet{doschek08} found a strong correlation between Doppler velocities of outflows and non-thermal velocities.  \citet{delzanna08} proposed chromospheric evaporation as a possible mechanism for the origin of the AR outflows.  Further, \citet{hara08} invoked `the hot plasma upflow near the base of the corona is direct evidence for impulsive heating'.  Reconnection over QSLs can naturally include these results.

We consider that there are at least four reconnection-related mechanisms that can drive AR outflows at QSLs:
\begin{enumerate}
\item impact of accelerated particles in denser lower layers leading to gentle chromospheric evaporation;
\item pressure gradient generated after the reconnection of two loops; 
\item small-scale reconnection jet-like outflows; and,
\item siphon flows along closed loops (see \S 4.6).
\end{enumerate}

We find that none of the mechanisms currently put forward in the literature is contradictory to reconnection occurring at QSLs as proposed by us, however, it is not the complete picture.  An additional outflow mechanism proposed by \citet{murray09} is not based on reconnection.  These authors suggest that continuous AR expansion compresses the neighboring magnetic field driving flows along `open' field or long loops.  These outflows appear at the boundary of ARs in the vicinity of QSLs, therefore, we suggest that AR outflows are caused by a combination of reconnection along QSLs and AR expansion in the vicinity of QSLs.

Finally, \citet{sakao07} and \citet{harra08} suggested that the AR outflows are a possible source of the slow solar wind.   \citet{liewer04} and \citet{ko06}, analyzing active region sources of the slow solar wind using $\it{in~situ}$ and remote sensing data, linked the active region sources of the solar wind to separatrices between loops connecting two different opposite polarity regions.  This is consistent with our suggestion of the relationship between QSLs and AR outflows.

\acknowledgements
We would like to thank the referee for his/her significant contribution which helped to greatly improve the paper.  Thanks to A. Wallace and G. Attrill for reading the revised manuscript.  Hinode is a Japanese mission developed and launched by ISAS/JAXA, collaborating with NAOJ as a domestic partner, NASA and STFC (UK) as international partners. Scientific operation of the Hinode mission is conducted by the Hinode science team organized at ISAS/JAXA. This team mainly consists of scientists from institutes in the partner countries. Support for the post-launch operation is provided by JAXA and NAOJ (Japan), STFC (U.K.), NASA (U.S.A.), ESA, and NSC (Norway).  DB thanks STFC for support via PhD studentship.  LvDG's work was partially supported by  the European Commission through the SOTERIA Network (EU FP7 Space Science Project No. 218816).  CHM acknowledges the Argentinean grants UBACyT X127 (UBA) and PICT 03-33370 (ANPCyT). CHM is a member of the Carrera del Investigador Cient\'\i fico (CONICET). The authors acknowledge financial support from ECOS-Sud (France) and MINCyT (Argentina) through their cooperative science program 
(N$^o$ A08U01).  MJM acknowledges financial assistance from STFC.

%Figure 1
\clearpage

%Figure 2
\clearpage

%Figure 3
\clearpage

%Figure 4
\clearpage

\clearpage

%Figure 6
\clearpage

%Figure 7
\clearpage

%Figure 5

\end{document}